\documentclass[cits]{PoS}

\usepackage{amsmath}
\usepackage{bm}
\usepackage[utf8x]{inputenc}
\usepackage{url}
\usepackage{subcaption}

\definecolor{red}{rgb}{0.76,0,0.16}
\definecolor{gray}{rgb}{0.5,0.5,0.5}
\definecolor{lightgray}{rgb}{0.7,0.7,0.7}
\definecolor{yellow}{cmyk}{0,0.2,1,0.1}
\definecolor{blue-dark}{cmyk}{0.8,0.4,0,0.4}
\definecolor{ucgray}{rgb}{0.02,0.22,0.31}
\definecolor{blue}{rgb}{0,0.396,0.741}
\definecolor{orange}{rgb}{1,0.4,0}
\definecolor{lapth-light-blue}{rgb}{0.4,0.64,0.85}
\definecolor{ivory}{rgb}{0.855,0.843,0.796}
\definecolor{green}{rgb}{0.569,0.675,0.420}
\definecolor{green-dark}{rgb}{0.286, 0.4761, 0.173}
\definecolor{red}{rgb}{0.768627, 0.027451, 0.105882}
\definecolor{purple}{rgb}{0.411765, 0.0313725, 0.352941}
\definecolor{lightgreen}{rgb}{0.710,0.792,0.510}

\usepackage{feynmp}
\makeatletter
\def\endfmffile{%
  \fmfcmd{\p@rcent\space the end.^^J%
          end.^^J%
          endinput;}%
  \if@fmfio
    \immediate\closeout\@outfmf
  \fi
  \ifnum\pdfshellescape=\@ne
    \immediate\write18{mpost \thefmffile}%
  \fi}
\makeatother

\title{Flavour anomalies and\\ (fundamental) partial compositeness}

\ShortTitle{Flavour anomalies and (fundamental) partial compositeness}

\author{\speaker{Peter Stangl}%
         \thanks{This article is based on~\cite{Niehoff:2015bfa} done in collaboration with Christoph Niehoff and David M. Straub, on~\cite{Sannino:2017utc} done in collaboration with Francesco Sannino, David M. Straub, and Anders Eller Thomsen, and on the PhD thesis of the author~\cite{Stangl:2018kty}.}
\\

        Laboratoire d’Annecy-le-Vieux de Physique Th\'eorique, UMR5108, Universit\'e de Savoie Mont-Blanc et CNRS, 9 Chemin de Bellevue, B.P. 110, F-74941, Annecy-le-Vieux Cedex, France\\
        E-mail: \email{peter.stangl@lapth.cnrs.fr}}

\abstract{Several measurements of B-meson decay observables show deviations from Standard Model (SM) predictions, some of them hinting at violation of lepton flavour universality (LFU).
I discusses how the anomalies in rare $B$ decays can be explained by partial compositeness. Partial compositeness is a key ingredient of models with a composite Higgs boson and generically leads to violation of LFU.
After presenting a simple model with partial compositeness that is able to explain the anomalies in rare $B$ decays, the flavour phenomenology of a minimal UV completion of a composite Higgs model with partial compositeness is discussed: the minimal fundamental partial compositeness (MFPC) model.
A virtue of the MFPC model is its capability of serving both as a solution to the naturalness problem of the SM and as an explanation of the flavour anomalies in rare $B$-meson decays.
In view of recent new measurements, the results on which this proceedings contribution is based are updated.
}

\FullConference{Corfu Summer Institute 2018 "School and Workshops on Elementary Particle Physics and Gravity"\\
		(CORFU2018)\\
		31 August - 28 September, 2018\\
		Corfu, Greece}

\begin{document}

\section{The flavour anomalies}\label{sec:anomalies}
Over the past few years, several measurements have shown evidence for deviations from the Standard Model (SM) predictions in $B$-meson decays, which are also known as the \emph{flavour anomalies}.
Some of these deviations hint at violation of lepton flavour universality (LFU), which is preserved by the SM to a high degree.
The measurements showing discrepancies can be divided into three categories:
\begin{enumerate}
 \item \label{anom:bsmumu} \textbf{The} $\mathbf{b\bm{\to} s\,\bm{\mu}^{\bm{+}} \bm{\mu}^{\bm{-}}}$ \textbf{anomaly}:
 Several measurements by the LHCb collaboration of observables based on the $b\to s\,\mu^+\mu^-$ transition deviate from the SM predictions by 2-3$\sigma$.
 These are in particular
 \begin{itemize}
  \item the angular observable $P_5'$ in $B\to K^*\mu^+\mu^-$~\cite{Aaij:2015oid},
  \item branching ratios of $B\to K\mu^+\mu^-$, $B\to K^{*}\mu^+\mu^-$, and $B_s\to \phi\mu^+\mu^-$~\cite{Aaij:2014pli,Aaij:2015esa,Aaij:2016flj}.
 \end{itemize}
 \item \label{anom:LFU_NC} \textbf{Hints for LFU violation in neutral current decays}:
 Measurements by the LHCb collaboration of the LFU ratios
 \begin{equation}
    R_{K^{(*)}}=\dfrac{BR(B\to K^{(*)}\mu^+\mu^-)}{BR(B\to K^{(*)}e^+e^-)},
 \end{equation}
 which are based on the neutral current transition $b\to s\,\ell^+\ell^-$, deviate from the SM prediction.
 In the SM, these ratios are unity to a very good approximation.
 $R_K$ measured in the $q^2$ bin\footnote{%
 $q^2$ is the dilepton invariant mass squared.} $[1.1,6.0]$~GeV$^2$ and $R_{K^*}$ measured in the bins $[0.045,1.1]$~GeV$^2$ and $[1.1,6.0]$~GeV$^2$ show deviations by about 2.5$\sigma$ each \cite{Aaij:2017vbb,Aaij:2019wad}.
 A recent measurement of $R_{K^*}$ in several $q^2$ bins by the Belle collaboration, which has considerably larger uncertainties than the LHCb measurement, is compatible with the LHCb result as well as with the SM prediction~\cite{RKstar_Belle_update}.
\item \label{anom:LFU_CC} \textbf{Hints for LFU violation in charged current decays}:
Measurements by BaBar, Belle, and LHCb of the LFU ratios
\begin{equation}
R_{D^{(*)}}=\dfrac{BR(B\to D^{(*)}\tau\nu)}{BR(B\to D^{(*)}\ell\nu)},
\quad\quad
\ell \in \{e,\mu\},
\end{equation}
which are based on the charged current transition $b\to c\,\ell\nu$, show a combined deviation from the SM predictions by $3.1\sigma$~\cite{Lees:2012xj,Lees:2013uzd,Huschle:2015rga,Sato:2016svk,Hirose:2016wfn,Aaij:2015yra,Aaij:2017uff,RDRDstar_Belle_preliminary,Abdesselam:2019dgh}.
\end{enumerate}
Interestingly, the $b\to s\,\mu^+\mu^-$ anomaly and the hints for LFU violation in $R_{K^{(*)}}$ can both be explained simultaneously by new physics (NP) contributing to the $b\to s\,\ell^+\ell^-$ transition
\cite{Altmannshofer:2017fio,Altmannshofer:2017yso,Ciuchini:2017mik,Capdevila:2017bsm,Geng:2017svp,DAmico:2017mtc,Alok:2017sui,Hurth:2017hxg,Alguero:2019ptt,Ciuchini:2019usw,Aebischer:2019mlg,Datta:2019zca,Kowalska:2019ley,Arbey:2019duh}.
At the $b$-quark-mass scale, such contributions can be described by an effective Hamiltonian
\begin{equation}\label{eq:Heff_bsll}
\mathcal H_\text{eff}^{bs\ell\ell}
= \mathcal H_\text{eff, SM}^{bs\ell\ell}
+ \mathcal H_\text{eff, NP}^{bs\ell\ell} \,,
\end{equation}
where $\mathcal H_\text{eff, NP}^{bs\ell\ell}$ parameterizes the NP effects and contains the terms\footnote{%
$\mathcal H_\text{eff, NP}^{bs\ell\ell}$ also contains terms involving e.g.\ the dipole operators $O^{(\prime) bs}_7$ and the scalar operators $O^{(\prime) \ell}_{S,P}$, which are not considered here.
For a recent analysis of NP effects in $b\to s\,\ell^+\ell^-$ at the $b$-quark scale taking into account also dipole and scalar operators, see e.g.~\cite{Aebischer:2019mlg}.
}
\begin{equation}
  \mathcal H_\text{eff, NP}^{bs\ell\ell}
  \supset -
  \frac{4G_F}{\sqrt{2}} V_{tb}V_{ts}^* \frac{e^2}{16\pi^2}
  \sum_{\ell=e,\mu}
  \sum_{i=9,10}
  \left(C^{\ell}_i O^{\ell}_i + C^{\prime \ell}_i O^{\prime \ell}_i\right)
  +
   \text{h.c.}\,,
\end{equation}
where the operators $O^{(\prime) \ell}_{9,10}$ are given by
\begin{align}
O_9^{\ell} &=
(\bar{s} \gamma_{\mu} P_{L} b)(\bar{\ell} \gamma^\mu \ell)\,,
&
O_9^{\prime \ell} &=
(\bar{s} \gamma_{\mu} P_{R} b)(\bar{\ell} \gamma^\mu \ell)\,,\label{eq:O9}
\\
O_{10}^{\ell} &=
(\bar{s} \gamma_{\mu} P_{L} b)( \bar{\ell} \gamma^\mu \gamma_5 \ell)\,,
&
O_{10}^{\prime \ell} &=
(\bar{s} \gamma_{\mu} P_{R} b)( \bar{\ell} \gamma^\mu \gamma_5 \ell)\,.\label{eq:O10}
\end{align}
Taking into account a large number of observables, several analyses have shown that a NP contribution to the muonic Wilson coefficients $C_9^{\mu}$ and $O_{10}^{\mu}$ can give a very good description of the experimental data on $b\to s\,\ell^+\ell^-$ processes~\cite{Altmannshofer:2017fio,Altmannshofer:2017yso,Ciuchini:2017mik,Capdevila:2017bsm,Geng:2017svp,DAmico:2017mtc,Alok:2017sui,Hurth:2017hxg,Alguero:2019ptt,Ciuchini:2019usw,Aebischer:2019mlg,Datta:2019zca,Kowalska:2019ley,Arbey:2019duh}.
In particular, the two scenarios with NP in a single combination of Wilson coefficients that yield the best fits to $b\to s\,\ell^+\ell^-$ data and can simultaneously explain the $b\to s\,\mu^+\mu^-$ anomaly and the $R_{K^{(*)}}$ measurements are~\cite{Aebischer:2019mlg}
\begin{itemize}
 \item $\mathbf{C_9^{\bm{\mu}}\bm{=-}0.95}^{\mathbf{+0.16}}_{\mathbf{-0.15}}$: a negative contribution to the coefficient of the operator that couples left-handed $b$ and $s$ quarks to a muon vector current,
 \item $\mathbf{C_9^{\bm{\mu}}\bm{=-}C_{10}^{\bm{\mu}}\bm{=-}0.53}^{\mathbf{+0.08}}_{\mathbf{-0.09}}$: a negative contribution to the coefficient of the operator that couples left-handed $b$ and $s$ quarks to left-handed muons.
\end{itemize}
Explaining only the $b\to s\,\mu^+\mu^-$ anomaly is also possible with lepton-universal Wilson coefficients. LFU violation in $R_{K^{(*)}}$ can also be explained using only electron Wilson coefficients.
However, the simultaneous explanation of both anomalies necessarily requires a contribution to the muon coefficients.

To address also the hints for LFU violation in the charged current $b\to c\,\ell\nu$ transitions, one has to consider additional effective operators not contained in eq.~(\ref{eq:Heff_bsll}).
In the SM, $b\to c\,\ell\nu$ decays are tree level processes, while the $b\to s\,\ell\ell$ decays are rare decays that are forbidden at tree level and are loop and CKM suppressed.
Therefore,
in order to generate the amount of LFU violation seen by experiments,
the NP contributions -- which have to compete with those from the SM -- need to be much larger for $b\to c\,\ell\nu$ than for $b\to s\,\ell\ell$.
Such large contributions require a relatively low NP scale and usually also affect other observables.
This makes it rather challenging to construct models that can explain $R_{D^{(*)}}$ while satisfying all other direct and indirect constraints.
In this article, the focus lies on solutions of the $b\to s\,\ell\ell$ anomalies. Still, LFU violation in charged current decays is considered in the context of the model discussed in section \ref{sec:MFPC}.

A plethora of models has been constructed to specifically address the flavour anomalies (for a review, see~\cite{Bifani:2018zmi}).
However, instead of building a new model, it is also interesting to investigate whether a model that has been constructed as solution to a problem of the SM can actually also explain the flavour anomalies.
One of the problems of the SM is the so called \emph{naturalness problem}.
This problem arises because the Higgs mass squared $m_h^2$ receives quantum correction to its bare mass squared $m_0^2$ that are proportional to the cutoff scale of the SM squared $\Lambda_{NP}^2$.
If there is no NP between the electroweak (EW) scale $\sim100$~GeV and the Planck scale $\sim10^{19}$~GeV, then the quantum corrections are of the size of the Planck scale squared and $m_0^2$ has to be extremely fine tuned to get a Higgs mass of the order of the EW scale.
To avoid this fine-tuning, a NP sector is required at a scale $\Lambda_{NP}$ not too far above the EW scale and this NP sector must not reintroduce the naturalness problem for energies higher than $\Lambda_{NP}$.
This is interesting since the flavour anomalies hint at NP degrees of freedom below a scale of $\Lambda_{NP}\sim100$~TeV~\cite{Altmannshofer:2017yso,DiLuzio:2017chi}, suggesting a possible connection to the NP sector that solves the naturalness problem.
Among the prime candidates for solutions to the naturalness problem are supersymmetric models like the MSSM and composite Higgs models (CHMs).
While it has been shown that it is not possible to accommodate the flavour anomalies in the MSSM~\cite{Altmannshofer:2014rta}, their explanation in CHMs is discussed here.

A key ingredient of many CHMs is a mixing between elementary and composite fermions that leads to the \emph{partial compositeness} of the SM-like mass eigenstates.
As has been shown in~\cite{Niehoff:2015bfa}, partial compositeness can play a prominent role in the explanation of the flavour anomalies in CHMs.
In section~\ref{sec:CHM}, a simple model featuring partial compositeness is presented as a solution of the $b\to s\,\ell\ell$ anomalies.
The flavour phenomenology of a UV completion of CHMs called fundamental partial compositeness is discussed in section~\ref{sec:MFPC} and it is shown that the minimal fundamental partial compositeness (MFPC) model is capable of explaining the anomalies in rare $B$ decays.

\section{Explaining the anomalies in rare B decays with partial compositeness}\label{sec:CHM}
CHMs can solve the naturalness problem of the SM since in these models, the Higgs field is not an elementary scalar but a bound state of a new strong interaction.
Therefore, above the compositeness scale $\Lambda_{NP}$ of this new strong interaction, there is simply no Higgs field and the naturalness problem does not arise.
Since no composite states have been observed so far except for the potentially composite Higgs, the scale $\Lambda_{NP}$ has to be considerably larger than the Higgs mass.
The lightness of the Higgs compared to the scale $\Lambda_{NP}$ can be explained if the Higgs is actually a pseudo-Nambu-Goldstone boson (pNGB) of a global symmetry that is spontaneously broken by the new strong interaction~\cite{Kaplan:1983fs,Dugan:1984hq}.

In this case, the Lagrangian of the model has the form
\begin{equation}\label{eq:L_CHM}
 \mathcal{L}_{CHM} =
 \textcolor{blue}{\mathcal{L}_{elemenary}} +
 \textcolor{green-dark}{\mathcal{L}_{composite}}
 + \textcolor{red}{\mathcal{L}_{mixing}},
\end{equation}
where $\textcolor{blue}{\mathcal{L}_{elemenary}}$ contains the SM fields except for the Higgs and $\textcolor{green-dark}{\mathcal{L}_{composite}}$ contains the composite Higgs as well as other composite bound states like composite vector bosons and composite fermions.
To generate the SM Yukawa couplings, the elementary fermions have to interact with the composite Higgs.
However, a direct interaction between the elementary fermions and the composite sector generically leads to experimentally excluded large contributions to flavour-changing neutral currents (FCNCs)~\cite{Eichten:1979ah,Dimopoulos:1980fj}.
An alternative proposed in~\cite{Kaplan:1991dc} is to
only couple the composite fermions directly to the Higgs
and to
introduce mixing terms
in $\textcolor{red}{\mathcal{L}_{mixing}}$ that linearly couple the elementary to the composite fermions.
\begin{figure}[b]
\center
$
\begin{fmffile}{PC}
\vcenter{\hbox{
\begin{fmfgraph*}(200,50)
\fmfset{arrow_len}{2mm}
\fmfset{thin}{1pt}
\fmfpen{thin}
\fmfstraight
\fmfleft{x,xp}
\fmfright{y,yp}
\fmf{plain,label=$\textcolor{blue}{f_L}$,f=(0,,0.396,,0.741)}{x,m1}
\fmf{double,label=$\textcolor{green-dark}{F_R}$,f=(0.408,,0.529,,0.235)}{m1,h}
\fmf{double,label=$\textcolor{green-dark}{F_L}$,f=(0.408,,0.529,,0.235)}{h,m2}
\fmf{plain,label=$\textcolor{blue}{f_R}$,f=(0,,0.396,,0.741)}{m2,y}
\fmf{phantom}{xp,z}
\fmf{phantom}{z,yp}
\fmf{dbl_dashes,label=$\textcolor{green-dark}{H}$,f=(0.408,,0.529,,0.235),tension=0}{h,z}
\fmfv{decor.shape=circle,decor.filled=full,decor.size=5,f=(0.408,,0.529,,0.235)}{h}
\fmfv{decor.shape=circle,decor.filled=empty,decor.size=5.5pt,label.dist=0,label=\textcolor{red}{+},f=(0.768627,,0.027451,,0.105882)}{m1,m2}
\end{fmfgraph*}}}
\end{fmffile}
$
\vskip20pt
\caption{Illustration of the Higgs coupling to composite fermions and the mixing between elementary and composite fermions.}
\label{fig:PC}
\end{figure}
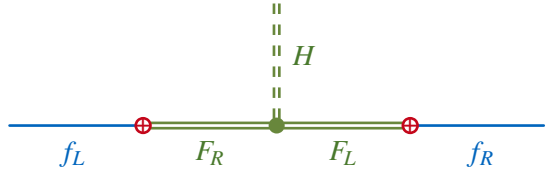
The mixing terms and the resulting coupling between the composite Higgs and the elementary fermions is shown in a diagrammatic way in fig.~\ref{fig:PC}.
It is interesting to note that in this setup, the coupling of the Higgs to the composite fermions can be chosen to be flavour universal and the complete flavour structure is then contained in the mixing terms.
Due to the mixing terms, the mass eigenstates in the model are actually superpositions of elementary and composite fields.
The light SM-like mass eigenstates are mainly elementary, but still \emph{partially composite}.
The size of the mixing terms determines the mixing angle between elementary states and SM-like mass eigenstates.
The sine of this mixing angle is usually called the \emph{degree of compositeness}.
The effective SM Yukawa couplings that couple the light mass eigenstates to the composite Higgs then depend on the product of a left-handed and a right-handed degree of compositeness.

\begin{figure}[t]
\centering
\begin{subfigure}[b]{0.49\textwidth}
\centering
$
\begin{fmffile}{bsll}
\vcenter{\hbox{
\begin{fmfgraph*}(135,105)
\fmfset{arrow_len}{2mm}
\fmfset{thin}{1pt}
\fmfpen{thin}
\fmfstraight

\fmfleft{x1,s,,,,,,,b,x2}
\fmfright{y1,mum,,,,,,,mup,y2}

\fmf{fermion,tension=2,f=(0,,0.396,,0.741)}{b,mb}
\fmf{double,tension=2,f=(0.408,,0.529,,0.235)}{mb,v1}
\fmf{fermion,tension=2,f=(0,,0.396,,0.741)}{ms,s}
\fmf{double,tension=2,f=(0.408,,0.529,,0.235)}{v1,ms}

\fmf{dbl_wiggly,tension=2/3,l.d=8,label=$\textcolor{green-dark}{Z'}$,f=(0.408,,0.529,,0.235)}{v1,v2}

\fmf{fermion,tension=2,f=(0,,0.396,,0.741)}{mup,mmup}
\fmf{double,tension=2,f=(0.408,,0.529,,0.235)}{v2,mmup}
\fmf{fermion,tension=2,f=(0,,0.396,,0.741)}{mmum,mum}
\fmf{double,tension=2,f=(0.408,,0.529,,0.235)}{mmum,v2}

\fmfv{decor.shape=circle,decor.filled=full,decor.size=5,f=(0.408,,0.529,,0.235)}{v1,v2}
\fmfv{decor.shape=circle,decor.filled=empty,decor.size=5.5pt,label.dist=0,label=\textcolor{red}{+},f=(0.768627,,0.027451,,0.105882)}{ms,mb,mmup,mmum}

\fmfv{lab=$\textcolor{blue}{b}$,l.angle=-180}{b}
\fmfv{lab=$\textcolor{blue}{s}$,l.angle=-180}{s}
\fmfv{lab=$\textcolor{blue}{\ell}$,l.angle=0}{mup}
\fmfv{lab=$\textcolor{blue}{\ell}$,l.angle=0}{mum}

\end{fmfgraph*}}}
\end{fmffile}
$
\vskip-10pt
\caption{}
\label{fig:LFUV:main_diags:bsll}
\end{subfigure}
\begin{subfigure}[b]{0.49\textwidth}
\centering
$
\begin{fmffile}{bsbs}
\vcenter{\hbox{
\begin{fmfgraph*}(135,105)
\fmfset{arrow_len}{2mm}
\fmfset{thin}{1pt}
\fmfpen{thin}
\fmfstraight

\fmfleft{x1,s,,,,,,,b,x2}
\fmfright{y1,mum,,,,,,,mup,y2}

\fmf{fermion,tension=2,f=(0,,0.396,,0.741)}{b,mb}
\fmf{double,tension=2,f=(0.408,,0.529,,0.235)}{mb,v1}
\fmf{fermion,tension=2,f=(0,,0.396,,0.741)}{ms,s}
\fmf{double,tension=2,f=(0.408,,0.529,,0.235)}{v1,ms}

\fmf{dbl_wiggly,tension=2/3,l.d=8,label=$\textcolor{green-dark}{Z'}$,f=(0.408,,0.529,,0.235)}{v1,v2}

\fmf{fermion,tension=2,f=(0,,0.396,,0.741)}{mup,mmup}
\fmf{double,tension=2,f=(0.408,,0.529,,0.235)}{v2,mmup}
\fmf{fermion,tension=2,f=(0,,0.396,,0.741)}{mmum,mum}
\fmf{double,tension=2,f=(0.408,,0.529,,0.235)}{mmum,v2}

\fmfv{decor.shape=circle,decor.filled=full,decor.size=5,f=(0.408,,0.529,,0.235)}{v1,v2}
\fmfv{decor.shape=circle,decor.filled=empty,decor.size=5.5pt,label.dist=0,label=\textcolor{red}{+},f=(0.768627,,0.027451,,0.105882)}{ms,mb,mmup,mmum}

\fmfv{lab=$\textcolor{blue}{b}$,l.angle=-180}{b}
\fmfv{lab=$\textcolor{blue}{s}$,l.angle=-180}{s}
\fmfv{lab=$\textcolor{blue}{b}$,l.angle=0}{mup}
\fmfv{lab=$\textcolor{blue}{s}$,l.angle=0}{mum}

\end{fmfgraph*}}}
\end{fmffile}
$
\vskip-10pt
\caption{}
\label{fig:LFUV:main_diags:bsbs}
\end{subfigure}
\caption{Tree-level contributions to
$b\to s\,\ell^+\ell^-$ (a) and $B_s$ mixing (b).
}\label{fig:LFUV:main_diags}
\end{figure}
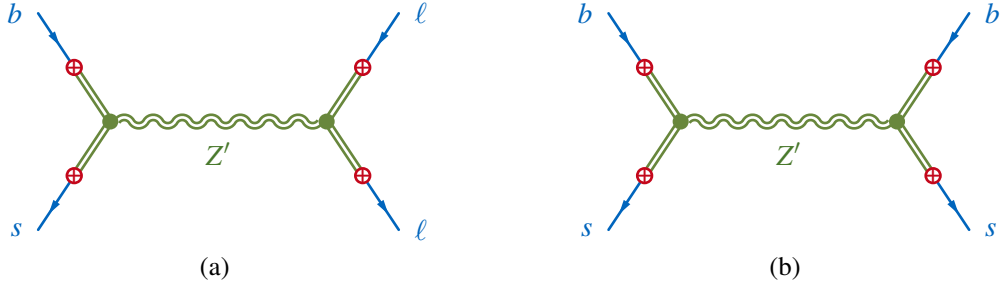

Given the above setup and considering only composite states with SM quantum numbers\footnote{%
An early explanation of the flavour anomalies in CHMs employs composite leptoquarks~\cite{Gripaios:2014tna}.
},  it has been shown in~\cite{Niehoff:2015bfa} that only the diagram in fig.~\ref{fig:LFUV:main_diags:bsll}, in which a neutral composite vector boson $Z'$ is exchanged, can result in an LFU violating tree-level contribution to $b\to s\,\ell^+\ell^-$.
This diagram contains a mixing of each external elementary fermion with its composite fermion partner.
This is necessary for a tree-level $Z'bs$ coupling as well as for a LFU violating $Z'\ell\ell$ coupling.
If muons have a sizable degree of compositeness while electrons have a negligible one, the diagram in fig.~\ref{fig:LFUV:main_diags:bsll} contributes essentially only to $b\to s\,\mu^+\mu^-$ and LFU is violated.
As discussed in section \ref{sec:anomalies}, the experimental data prefers either
\begin{itemize}
 \item a negative contribution to $C_9^{\mu}$ or
 \item a negative contribution to $C_9^{\mu}=-C_{10}^{\mu}$.
\end{itemize}
Both solutions require left-handed $b$ and $s$ quarks with a sizable degrees of compositeness.
The first solution involves a muon vector current and would thus require sizable degrees of compositeness of left-handed and right-handed muons.
However, the product of both enters the effective SM Yukawa coupling of the muon and has to be small to yield the correct muon mass.
The second solution, on the other hand, requires only a sizable degree of compositeness of left-handed muons $s_{\mu_L}$. This second solution seems possible.
Several experimental bounds apply to this scenario:
\begin{itemize}
 \item $\mathbf{B_s}$-$\mathbf{\bar{B}_s}$ \textbf{mixing}: Given the flavour changing couplings of left-handed $b$ and $s$ quarks to the $Z'$, a NP contribution to $B_s$-$\bar{B}_s$ mixing from the diagram in fig.~\ref{fig:LFUV:main_diags:bsbs} cannot be avoided.
Hence, the experimental bound from $B_s$-$\bar{B}_s$ mixing provides an upper bound on the size of the $Z'bs$ coupling and in turn leads to a \textbf{lower bound on $\mathbf{s_{\bm{\mu}_L}}$}.
Using the diagram in fig.~\ref{fig:LFUV:main_diags:bsbs}, one can express the $Z'bs$ coupling in terms of a shift in the mass difference $\Delta M_s$ in $B_s$ mixing.
\item \textbf{Electroweak precision tests (EWPTs)}: The presence of a non-zero $Z'\mu_L\mu_L$ coupling generically implies corrections to the $Z\mu\mu$, $W\mu\nu_\mu$, and $Z\nu_\mu\nu_\mu$ couplings, which are constrained by EWPTs.
\begin{figure}[t]
\center
 \includegraphics[width=\textwidth]{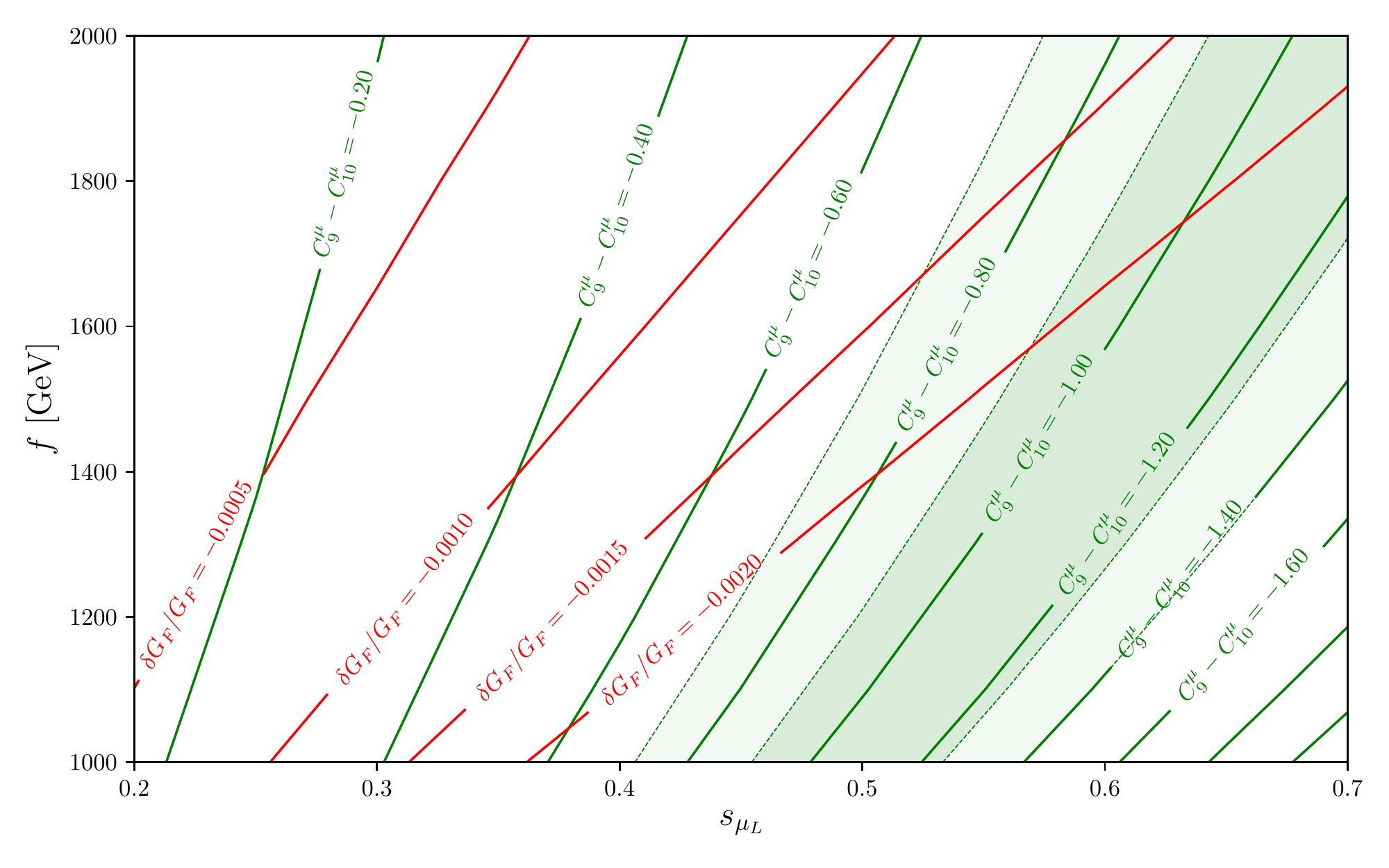}
 \caption{Contours of constant $C_9^{\mu}-C_{10}^{\mu}$ when assuming a 10\% shift in $\Delta M_s$ (green lines) and contours of constant relative shift in the Fermi constant $\delta G_F/G_F$ (red lines).
 The shaded areas correspond to the 1$\sigma$ (dark green) and 2$\sigma$ (light green) regions around the best-fit value of a global fit~\cite{Aebischer:2019mlg} in a scenario with a NP contribution to $C_9^{\mu}=-C_{10}^{\mu}$.
 }\label{fig:comp_mu_result}
\end{figure}
\begin{itemize}
 \item $\mathbf{Z\bm{\mu\mu}}$: Constraints from a modification of this coupling can be avoided by protecting\footnote{%
 Only the tree-level coupling can be protected and one-loop corrections might be relevant~\cite{Grojean:2013qca}. However, they depend on the particle content of a complete model and cannot be computed reliably in the simplified model considered here.
 } it with a discrete symmetry~\cite{Agashe:2006at,Agashe:2009tu}.
 \item $\mathbf{W\bm{\mu\nu_\mu}}$: The modification of this coupling due to the presence of a non-zero $s_{\mu_L}$ results in a NP contribution to the muon lifetime.
 This shifts the Fermi constant $G_F$, which is extracted from muon decay.
 The constraint on $G_F$ is strongly correlated with the constraint on the electroweak $T$ parameter such that the bound on the latter can be translated into a bound on the relative shift in the Fermi constant $|\delta G_F/G_F|\lesssim 0.002$.
 This then leads to an \textbf{upper bound on $\mathbf{s_{\bm{\mu}_L}}$}.
 \item $\mathbf{Z\bm{\nu_\mu\nu_\mu}}$: The contribution to this coupling is equal\footnote{This equality holds at tree-level if the $Z\mu\mu$ coupling is protected by a discrete symmetry.} to the contribution to the $W\mu\nu_\mu$ coupling and leads to a shift in the effective number of light neutrino species $N_\nu$.
 Interestingly, this shift does not lead to a constraint but actually improves the agreement with LEP data~\cite{ALEPH:2005ab}, which shows a 2$\sigma$ deviation from the SM value of $N_\nu$.
\end{itemize}
\end{itemize}

Expressing the $Z'bs$ coupling in terms of a shift in $\Delta M_s$, the diagram in fig.~\ref{fig:LFUV:main_diags:bsll} leads to a contribution to $C_9^{\mu}-C_{10}^{\mu}$ given by
\begin{equation}
C_9^\mu-C_{10}^\mu \approx
 \pm0.92
\left[\frac{1.7\,\text{TeV}}{f}\right]
\left[\frac{s_{\mu_L}}{0.6}\right]^2
\left[\frac{|\Delta M_s-\Delta M_s^\text{SM}|}{0.1\,\Delta
M_s^\text{SM}}\right]^{1/2}
\,,
\end{equation}
where $f$ is the NP scale\footnote{%
$f$ sets the mass scale of the composite resonances. The mass of the composite vector boson $Z'$ is given by $m_{Z'}^2=\frac{f^2\,g_{Z'}^2}{2}$, where $g_{Z'}$ is the coupling constant associated with the coupling between the $Z'$ and the composite fermions.
In a model in which the composite Higgs is a pNGB, $f$ is the pNGB decay constant (in analogy with the pion decay constant in QCD).
}.
Saturating the bound from $B_s$-$\bar{B}_s$ mixing by assuming a 10\% correction to $\Delta M_s$, the result only depends on the degree of compositeness of the left-handed muon $s_{\mu_L}$ and on the NP scale $f$. Lines of constant $C_9^{\mu}-C_{10}^{\mu}$ using this assumption are shown in fig.~\ref{fig:comp_mu_result} along with lines of constant shifts of the Fermi constant.
Requiring $|\delta G_F/G_F|\lesssim 0.002$, one finds that the $b\to s\,\mu^+\mu^-$ anomaly and the hints for LFU violation in neutral current decays can be explained with a sizable degree of compositeness of left-handed muons $s_{\mu_L}\approx0.6$ and a NP scale $f\approx1.7$ TeV.

\section{Flavour physics and flavour anomalies in the MFPC model}\label{sec:MFPC}

The analysis in section~\ref{sec:CHM} shows that a simplified model with partial compositeness can explain the anomalies in rare $B$ decays.
In view of this proof of principle, it is interesting to investigate whether such an explanation is still possible in
a UV-complete model.
UV complete fundamental partial compositeness (FPC) models have been proposed in~\cite{Sannino:2016sfx}\footnote{%
For other attempts to construct UV-complete CHMs, see~\cite{Caracciolo:2012je,Barnard:2013zea,Ferretti:2013kya,Ferretti:2014qta,Vecchi:2015fma}.
}.
In these models, a new strong force called technicolor (TC) confines new TC-charged fermions and scalars.
These so called technifermions $\mathcal{F}$ and techniscalars $\mathcal{S}$ are supposed to form bound states below the confinement scale~$\Lambda_\text{TC}$.
In particular, a composite Higgs boson can be realised as an $(\mathcal{F}\mathcal{F})$ bound state, whereas $(\mathcal{F}\mathcal{S})$ bound states provide composite fermions that can mix with elementary fermions $f$ and thus lead to partial compositeness.
The Lagrangian of an FPC model takes the form
\begin{equation}\label{eq:L_FPC}
 \mathcal{L}_{FPC} =
 \textcolor{blue}{\mathcal{L}_{elemenary}} +
 \textcolor{green-dark}{\mathcal{L}_{TC}}
 + \textcolor{red}{\mathcal{L}_{Yukawa}},
\end{equation}
where $\textcolor{blue}{\mathcal{L}_{elemenary}}$ contains the SM fields except for the Higgs and $\textcolor{green-dark}{\mathcal{L}_{TC}}$ contains the TC gauge bosons, the technifermions, and the techniscalars.
Partial compositeness requires that $(\mathcal{F}\mathcal{S})$ bound states have the same quantum numbers as the SM fermions.
In this case, the symmetries of the FPC model allow Yukawa coupling terms that couple an elementary fermion to a technifermion and a techniscalar.
These terms are contained in $\textcolor{red}{\mathcal{L}_{Yukawa}}$.

It is instructive to compare the FPC model in the UV above $\Lambda_\text{TC}$, which is described by the Lagrangian in eq.~(\ref{eq:L_FPC}), to its effective theory in the IR below $\Lambda_\text{TC}$, which is a CHM described by the Lagrangian in eq.~(\ref{eq:L_CHM}).
When going from the UV to the IR the different parts of the Lagrangian are replaced as follows:
\[
\begin{aligned}
\textcolor{blue}{\mathcal{L}_{elemenary}}
&\quad\xrightarrow{\text{IR}}\quad
\textcolor{blue}{\mathcal{L}_{elemenary}},\\
\textcolor{green-dark}{\mathcal{L}_{TC}}
&\quad\xrightarrow{\text{IR}}\quad
\textcolor{green-dark}{\mathcal{L}_{composite}},\\
\textcolor{red}{\mathcal{L}_{Yukawa}}
&\quad\xrightarrow{\text{IR}}\quad
\textcolor{red}{\mathcal{L}_{mixing}},
\end{aligned}
\]
i.e. $\textcolor{blue}{\mathcal{L}_{elemenary}}$ is unchanged while $\textcolor{green-dark}{\mathcal{L}_{TC}}$ and $\textcolor{red}{\mathcal{L}_{Yukawa}}$ in the FPC model in the UV are replaced by $\textcolor{green-dark}{\mathcal{L}_{composite}}$ and $\textcolor{red}{\mathcal{L}_{mixing}}$ in the CHM in the IR.
The fact that the Yukawa terms correspond to the mixing terms can also be illustrated in a diagrammatic way,
\[
\begin{fmffile}{fund_yuk}
\vcenter{\hbox{
\begin{fmfgraph*}(50,5)
\fmfset{arrow_len}{2mm}
\fmfset{thin}{1pt}
\fmfpen{thin}
\fmfstraight
\fmfleft{f}
\fmfright{F,S}
\fmf{plain,label=$\textcolor{blue}{f}$,tension=2.7,f=(0,,0.396,,0.741)}{f,y}
\fmf{plain,label=$\textcolor{green-dark}{\mathcal{F}}$,f=(0.408,,0.529,,0.235),right=.07}{y,F}
\fmf{dashes,label=$\textcolor{green-dark}{\mathcal{S}}$,f=(0.408,,0.529,,0.235),right=.07}{S,y}
\fmfv{decor.shape=circle,decor.filled=full,decor.size=5.5pt,f=(0.768627,,0.027451,,0.105882)}{y}
\end{fmfgraph*}}}
\end{fmffile}
\quad\xrightarrow{\text{IR}}\quad
\begin{fmffile}{mix}
\vcenter{\hbox{
\begin{fmfgraph*}(50,20)
\fmfset{arrow_len}{2mm}
\fmfset{thin}{1pt}
\fmfpen{thin}
\fmfstraight
\fmfleft{f}
\fmfright{F}
\fmf{plain,label=$\textcolor{blue}{f}$,f=(0,,0.396,,0.741)}{f,y}
\fmf{double,label=$\textcolor{green-dark}{F}$,f=(0.408,,0.529,,0.235)}{y,F}
\fmfv{decor.shape=circle,decor.filled=empty,decor.size=5.5pt,label.dist=0,label=\textcolor{red}{+},f=(0.768627,,0.027451,,0.105882)}{y}
\end{fmfgraph*}}}
\end{fmffile},
\]
\vspace{-10pt}\newline
i.e. the Yukawa terms coupling $f$, $\mathcal{F}$, and $\mathcal{S}$ become mixing terms once the fermionic bound states ${F \sim (\mathcal{F}\mathcal{S})}$ are formed.

For investigating the phenomenology of an FPC model at and below the EW scale, it is convenient to consider an effective field theory (EFT) in which only the elementary fields and the pNGBs are kept as dynamical degrees of freedom, while all other effects of the strongly coupled TC sector are encoded in Wilson coefficients of effective operators.
For the minimal\footnote{%
The MFPC model is minimal in the sense that its pNGB sector features only one singlet scalar in addition to the electroweak doublet that serves as the composite Higgs field.
} FPC (MFPC) model, this EFT has been constructed in~\cite{Cacciapaglia:2017cdi}.
In the MFPC model, the flavour structure is completely determined by the fundamental Yukawa terms that couple elementary fermions to technifermions and techniscalars.
In the MFPC-EFT, these fundamental Yukawa couplings enter the Wilson coefficients of effective Operators.
In particular, the Wilson coefficients that yield the mass terms of SM fermions and the CKM matrix are proportional to a product of two fundamental Yukawa coupling Matrices, while the coefficients of four-fermion Operators are proportional to a product of four fundamental Yukawa coupling matrices.
The latter originate e.g. from box-like diagrams in the UV, as depicted in the following:
\vskip15pt
 \[
\begin{fmffile}{FPC_box}
\vcenter{\hbox{
\begin{fmfgraph*}(50,50)
\fmfset{arrow_len}{2mm}
\fmfset{thin}{.7pt}
\fmfpen{thin}
\fmfstraight
\fmfleft{f3,f2}
\fmfright{f4,f1}
%
\fmf{plain,tension=3,f=(0,,0.396,,0.741)}{f1,x1}
\fmf{plain,tension=3,f=(0,,0.396,,0.741)}{f2,x2}
\fmf{plain,tension=3,f=(0,,0.396,,0.741)}{f3,x3}
\fmf{plain,tension=3,f=(0,,0.396,,0.741)}{f4,x4}
\fmf{plain,label=$\textcolor{green-dark}{\mathcal{F}}$,f=(0.408,,0.529,,0.235),right=0}{x3,x2}
\fmf{plain,label=$\textcolor{green-dark}{\mathcal{F}}$,f=(0.408,,0.529,,0.235),right=0}{x4,x1}
\fmf{dashes,label=$\textcolor{green-dark}{\mathcal{S}}$,f=(0.408,,0.529,,0.235),right=0}{x2,x1}
\fmf{dashes,label=$\textcolor{green-dark}{\mathcal{S}}$,f=(0.408,,0.529,,0.235),right=0}{x3,x4}
\fmfv{decor.shape=circle,decor.filled=full,decor.size=4.2pt,f=(0.768627,,0.027451,,0.105882)}{x1,x2,x3,x4}
\fmfv{label=$\textcolor{blue}{{f_1}_i}$}{f1}
\fmfv{label=$\textcolor{blue}{{f_2}_j}$}{f2}
\fmfv{label=$\textcolor{blue}{{f_3}_k}$}{f3}
\fmfv{label=$\textcolor{blue}{{f_4}_l}$}{f4}
\end{fmfgraph*}}}
\end{fmffile}
\hskip30pt
\xrightarrow{\text{EFT}}
\hskip30pt
\begin{fmffile}{FPC_blob}
\vcenter{\hbox{
\begin{fmfgraph*}(50,50)
\fmfset{arrow_len}{2mm}
\fmfset{thin}{.7pt}
\fmfpen{thin}
\fmfstraight
\fmfleft{f3,f2}
\fmfright{f4,f1}
\fmf{plain,tension=3,f=(0,,0.396,,0.741)}{f1,x1}
\fmf{plain,tension=3,f=(0,,0.396,,0.741)}{f2,x2}
\fmf{plain,tension=3,f=(0,,0.396,,0.741)}{f3,x3}
\fmf{plain,tension=3,f=(0,,0.396,,0.741)}{f4,x4}
\fmf{phantom,tension=12}{x1,x}
\fmf{phantom,tension=12}{x2,x}
\fmf{phantom,tension=12}{x3,x}
\fmf{phantom,tension=12}{x4,x}
\fmfv{decor.shape=circle,decor.filled=shaded,decor.size=13pt,f=(0.408,,0.529,,0.235)}{x}
\fmfv{decor.shape=circle,decor.filled=full,decor.size=4.2pt,f=(0.768627,,0.027451,,0.105882)}{x1,x2,x3,x4}
\fmfv{label=$\textcolor{blue}{{f_1}_i}$}{f1}
\fmfv{label=$\textcolor{blue}{{f_2}_j}$}{f2}
\fmfv{label=$\textcolor{blue}{{f_3}_k}$}{f3}
\fmfv{label=$\textcolor{blue}{{f_4}_l}$}{f4}
\end{fmfgraph*}}}
\end{fmffile}
\]
\vskip17pt
\[
\propto (\textcolor{red}{y_{f_1}^T y_{f_2}})_{ij}\,(\textcolor{red}{y_{f_3}^T y_{f_4}})_{kl}
\]
\vspace{-12pt}\newline
Although the overall size of the Wilson coefficients depends on non-perturbative effects from the strong TC interaction between $\mathcal{F}$ and $\mathcal{S}$, the flavour structure of the Wilson coefficients is completely determined by the fundamental Yukawa couplings and therefore independent of the strong interaction effects.
This makes the MFPC-EFT especially suitable for an analysis of the flavour phenomenology of the MFPC model.

Such an analysis has been performed in~\cite{Sannino:2017utc} and is described in the following.
It considers all 37 parameters of the MFPC-EFT that are relevant for the flavour observables.
These are 14 parameters encoding the strong interaction effects, 22 parameters that specify the fundamental Yukawa matrices, and $\Lambda_{TC}$ that sets the scale of the strong coupling.
These parameters are then constrained by requiring correct SM fermion masses and CKM elements and by considering constraints from electroweak scale observables and low-energy flavour observables.
The constrained parameter space is then used to make predictions for the observables related to the flavour anomalies, in particular the LFU observables $R_{K^{(*)}}$ and $R_{D^{(*)}}$.

At the EW scale, constraints from the LEP measurements~\cite{ALEPH:2005ab} of the $\mathbf{Z}$ \textbf{partial widths} are considered.
These observables are defined as
\[
 R_e=\frac{\Gamma(Z\to q\bar{q})}{\Gamma(Z\to e\bar{e})},
 \quad\quad
 R_\mu=\frac{\Gamma(Z\to q\bar{q})}{\Gamma(Z\to \mu\bar{\mu})},
 \quad\quad
 R_\tau=\frac{\Gamma(Z\to q\bar{q})}{\Gamma(Z\to \tau\bar{\tau})},
\]
\[
 R_b=\frac{\Gamma(Z\to b\bar{b})}{\Gamma(Z\to q\bar{q})},
 \quad\quad
 R_c=\frac{\Gamma(Z\to c\bar{c})}{\Gamma(Z\to q\bar{q})}.
\]
Their theoretical predictions are computed in the MFPC-EFT at the EW scale.

For taking into account the low-energy flavour observables, the MFPC-EFT is first matched to the weak effective theory (WET) at the EW scale by integrating out all particles heavier than the $b$ quark.
Theoretical predictions for all flavour observables are then computed from Wilson coefficients in the WET.
The flavour observables considered as constraints are
\begin{itemize}
 \item Meson-antimeson mixing observables:
 \begin{itemize}
 \item Indirect $CP$ violation in
 kaon
 mixing: $\bm{\epsilon_K$}~\cite{Agashe:2014kda}.
 \item Mixing-induced $CP$ asymmetries in the $B_d$ and $B_s$ systems: $\bm{S_{\bm{\psi K_S}}}$ and $\bm{S_{\bm{\psi \phi}}}$~\cite{Amhis:2014hma}.
 \item Mass differences in $B_d$ and $B_s$ systems: $\bm{\Delta M_{d}}$ and $\bm{\Delta M_{s}}$~\cite{Amhis:2014hma}.
 \end{itemize}
 \item Charged-current semi-leptonic decays:
 \begin{itemize}
 \item $\textbf{BR}\bm{(\pi^{\bm{+}}\to e\nu)}$~\cite{Aguilar-Arevalo:2015cdf}, based on $d \to ue\nu$.
 \item $\textbf{BR}\bm{(K^{\bm{+}}\to\mu\nu)}$, $\textbf{BR}\bm{(K^{\bm{+}}\to e\nu)}/\textbf{BR}\bm{(K^{\bm{+}}\to\mu\nu)}$~\cite{Agashe:2014kda}, based on $s\!\to\! ue\nu$, $s\!\to\! u\mu\nu$.
 \item $\textbf{BR}\bm{(B\to De\nu)}$, $\textbf{BR}\bm{(B\to D\mu\nu)}$~\cite{Amhis:2014hma}, based on $b\to ce\nu$, $b\to c\mu\nu$.
 \end{itemize}
 These observables are especially important to constrain the CKM elements and to provide tests of $e$-$\mu$ universality.
\end{itemize}

A parameter scan of the MFPC-EFT is challenging due to the large number of parameters.
Therefore, a numerical method is employed that consists of two steps.

\begin{itemize}
\item Step 1:
\begin{itemize}
 \item For applying the constraints from quark masses and CKM elements, the $\chi^2$ function\footnote{%
 The $\chi^2$ function has a small value if the theoretical predictions are close to the experimental observations and therefore can be used to quantify the agreement between theoretical predictions and experimental data.
 For more details, see~\cite{Sannino:2017utc,Stangl:2018kty}.
 } $\chi^2_{\mathrm mass,CKM}$ is constructed, which depends on only 19 of the 37 parameters.
 \item A numerical minimization of $\chi^2_{\mathrm mass,CKM}$ is performed for 100\,k random starting points.
 \item The regions around the local minima of $\chi^2_{\mathrm mass,CKM}$ are sampled using Markov Chains from the \texttt{pypmc} package~\cite{pypmc}.
 For each local minimum, 1000 points close to the minimum are determined.
\end{itemize}
This first step yields 100\,M parameter points that predict the correct quark masses and CKM elements.
\item Step 2:
\begin{itemize}
 \item The remaining 18 parameters are chosen randomly for each of the 100\,M points that have been determined in step 1.
 \item For each of the 100\,M parameters points, predictions for the EW scale and flavour observables listed above are computed using the \texttt{flavio} code~\cite{Straub:2018kue}.
 \item The predictions are compared to experimental data.
 Parameter points that are excluded by experimental data are discarded.
\end{itemize}
\end{itemize}
This two-step numerical parameter scan results in parameter points that predict the correct quark masses and CKM elements and satisfy all applied constraints from EW scale and flavour observables.
As expected from a comparison with other analyses of models with partial compositeness~\cite{Csaki:2008zd,Blanke:2008zb,Bauer:2009cf}, the observable $\epsilon_K$, which measures indirect $CP$ violation in kaon mixing, provides the strongest constraint.
However, still a large number of points is found that can satisfy all constraints.
These parameter points are then used to make predictions for the LFU observables $R_{D^{(*)}}$ and $R_{K^{(*)}}$

\begin{figure}[t]
\center
 \includegraphics[width=0.5\textwidth]{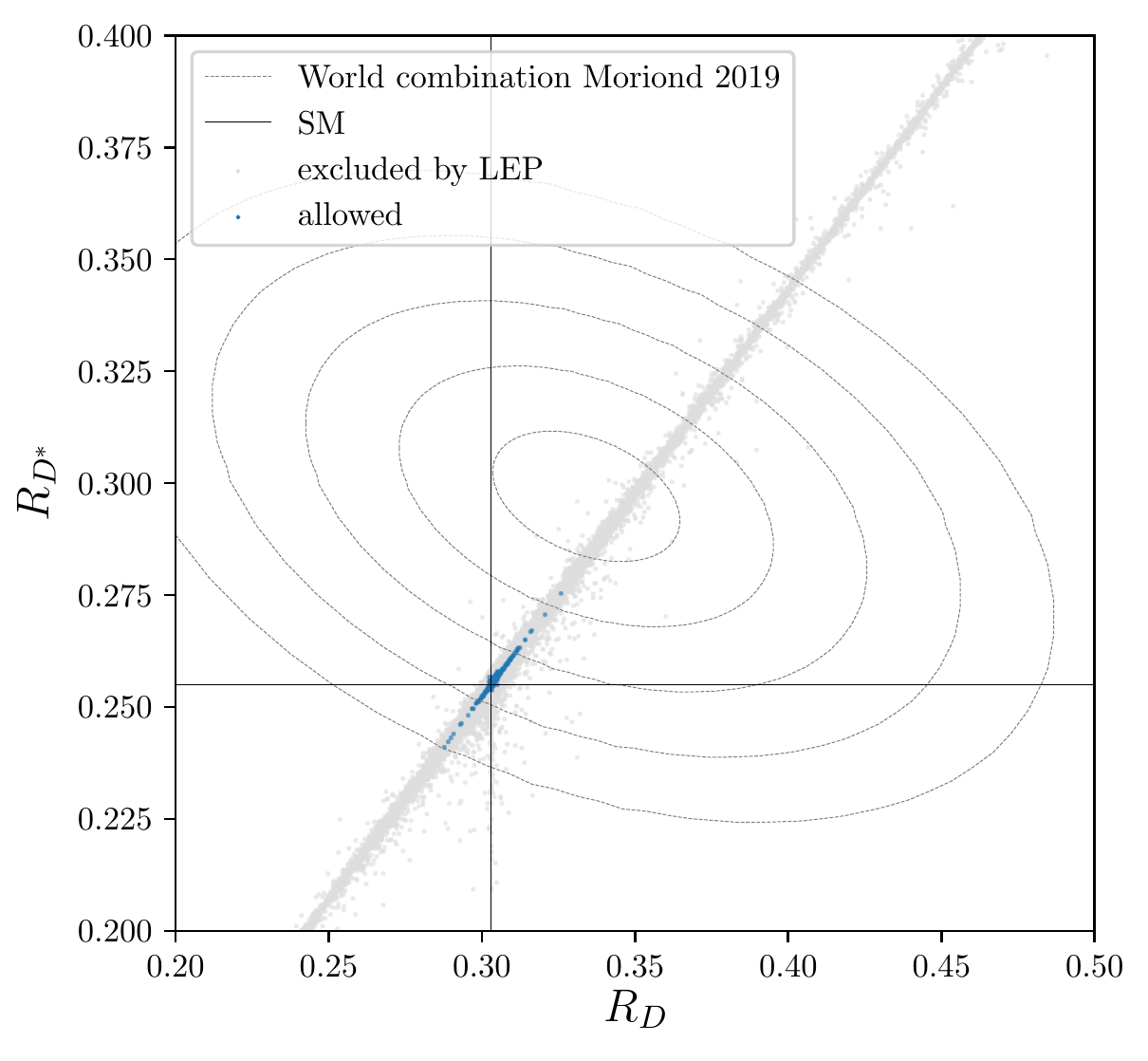}
 \caption{Predictions for LFU observables $R_{D^{(*)}}$.
 Gray points are excluded by LEP data. Blue points are allowed by all constraints.
 The plot also shows the 1-5$\sigma$ contours of the experimental world combination as presented at Moriond 2019~\cite{RDRDstar_Belle_preliminary}.
 }\label{fig:RD}
\end{figure}
The predictions for $R_{D^{(*)}}$ are shown in figure~\ref{fig:RD}.
It is found that the model cannot accommodate the central value of the Moriond 2019 world combination~\cite{RDRDstar_Belle_preliminary}.
Such large values for $R_{D}$ and $R_{D^{*}}$ would require a huge degree of compositeness of the $\tau$ lepton, which is excluded by the LEP measurements of $Z$ partial widths.
However, the tension with experimental data can be slightly reduced compared to the SM.

\begin{figure}[t]
\center
 \includegraphics[width=0.49\textwidth]{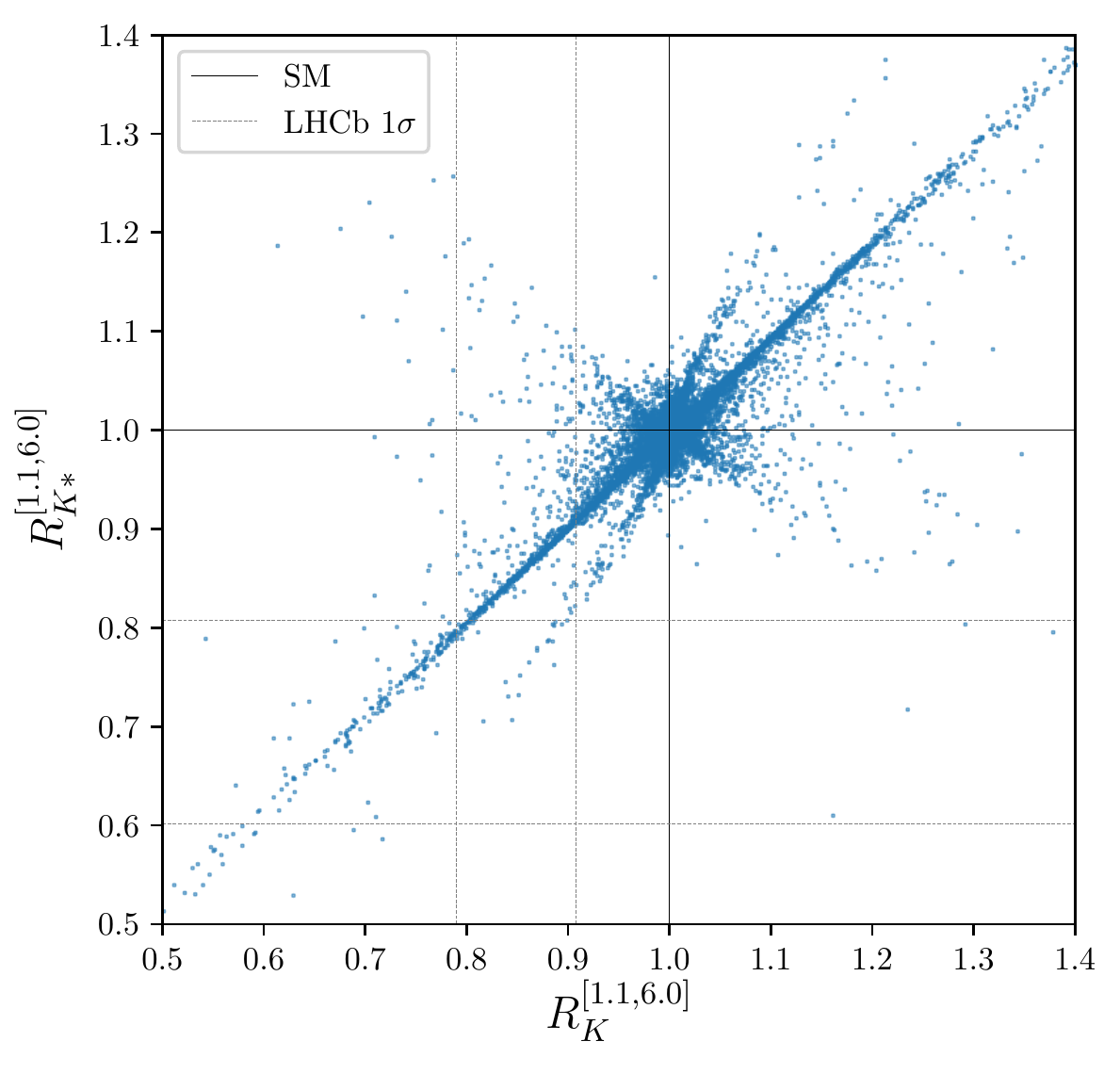}
 \includegraphics[width=0.49\textwidth]{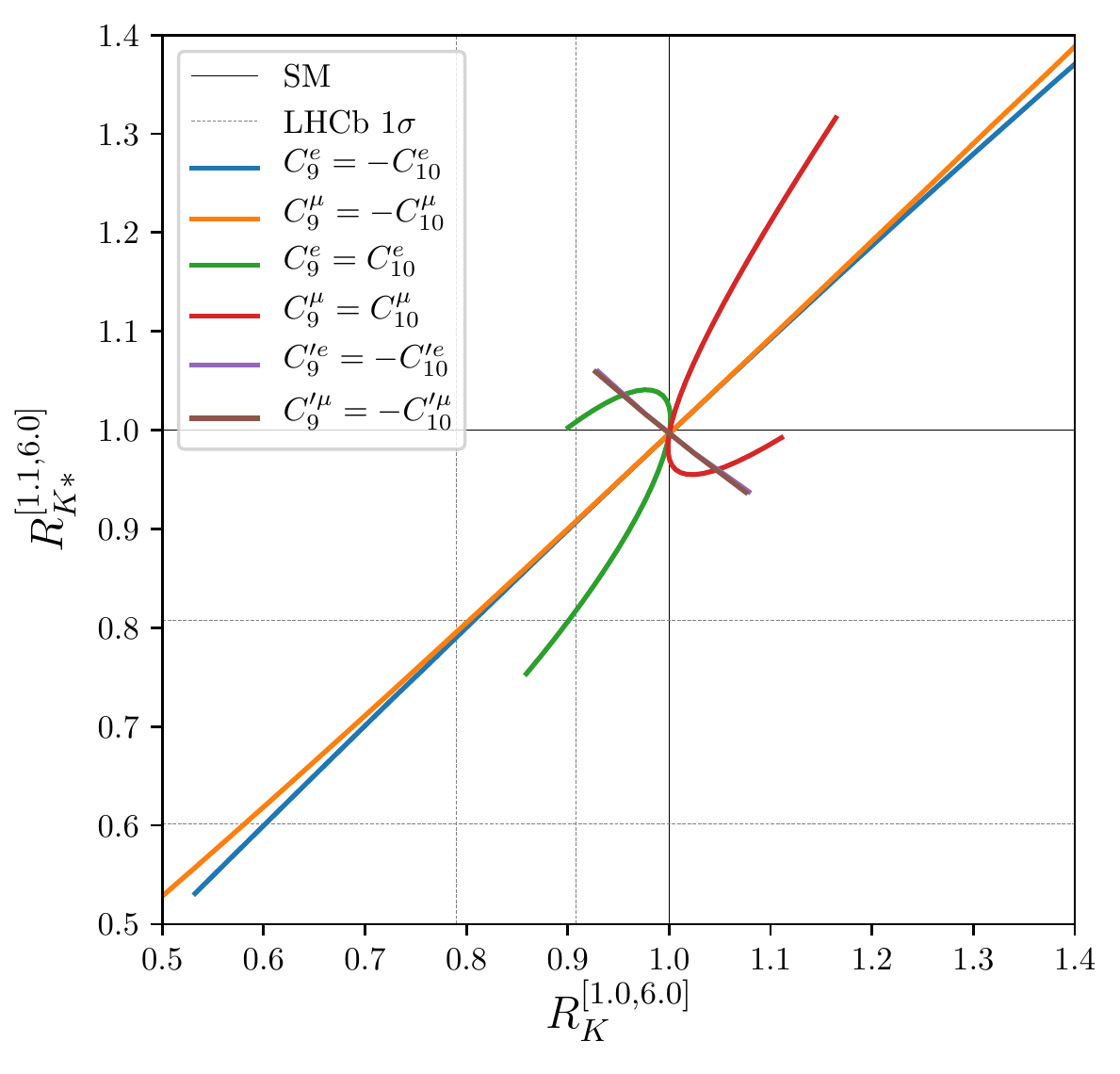}
 \caption{Left: Predictions for LFU observables $R_{K^{(*)}}$ in the MFPC model for points allowed by all constraints.
 Right:
 Generic predictions in several NP scenarios. Unprimed Wilson coefficients are varied between $-1.5$ and $1.5$; primed coefficients are varied between $-0.15$ and $0.15$.
 }\label{fig:RK}
\end{figure}
Predictions for the LFU observables $R_{K}$ and $R_{K^{*}}$ in the MFPC model for all allowed points are shown in figure~\ref{fig:RK} left.
These points roughly correspond to the generic NP scenarios depicted in figure~\ref{fig:RK} right.
The plots show that the MFPC model is actually able to explain the hints for LFU violation in neutral-current $B$ decays at the 1$\sigma$ level.
Such an explanation corresponds to a NP contribution to either $C_9^{\mu}=-C_{10}^{\mu}<0$, $C_9^{e}=-C_{10}^{e}>0$, or $C_9^{e}=C_{10}^{e}<0$.
While the scenarios featuring NP contributions to the electron Wilson coefficients can explain $R_{K^{(*)}}$, an explanation of the $b\to s\mu\mu$ anomaly requires a contribution to muon Wilson coefficients, singling out the scenario $C_9^{\mu}=-C_{10}^{\mu}<0$.
The MFPC model is thus capable of explaining all anomalies in rare $B$ decays.
\section{Conclusions}
Tensions between experimental data and SM predictions of $B$-decay observables hint at LFU violating NP. Models with partial compositeness generically violate LFU and are thus good candidates for explaining these tensions.
Using most recent experimental data, the analyses originally performed in~\cite{Niehoff:2015bfa,Sannino:2017utc,Stangl:2018kty} are updated.
It is found that
\begin{itemize}
 \item A simplified model with partial compositeness can explain the anomalies in rare $B$ decays.
 \begin{itemize}
  \item The explanation corresponds to a NP contribution to $C_{9}^{\mu}=-C_{10}^{\mu}<0$.
  \item A deviation in $B_{s}$-$\bar{B}_{s}$ mixing is predicted.
  \item A correction to the Fermi constant yields the strongest constraint from EWPTs if tree level $Z$ couplings are protected by a discrete symmetry.
 \end{itemize}
 \item A comprehensive numerical analysis of flavour and EW scale effects of a UV complete model featuring partial compositeness, the MFPC model, has been performed.
  \begin{itemize}
  \item The numerical method used in this analysis makes it possible to perform a scan of the high dimensional parameter space.
   \item The strongest constraint is due to $\epsilon_K$, but a large number of parameter points satisfy this constraint.
   \item Very large deviations of $R_{D^{(*)}}$ from their SM values are in conflict with LEP measurements of Z partial widths such that the central values of the Moriond 2019 experimental world combination cannot be accommodated by the model. However, the tensions can be slightly reduced.
   \item All anomalies in rare B decays can be explained by the MFPC model.
   This explanation corresponds to a NP contribution to $C_9^{\mu}=-C_{10}^{\mu}<0$.
  \end{itemize}

\end{itemize}

\section*{Acknowledgements}
The author thanks Christoph Niehoff, Francesco Sannino, David M. Straub, and Anders Eller Thomsen for the collaborations this article is based on and the organizers of the Corfu Summer Institute, in particular George Zoupanos, for the invitation to Corfu and the pleasant workshop.

\bibliographystyle{JHEP}
\bibliography{bibliography}

\end{document}